# The development of IBIC microscopy at the 100 kV ion implanter of the University of Torino (LIUTo) and the application for the assessment of the radiation hardness of a silicon photodiode.

Emilio Corte [1,2,#], Alberto Bortone [2,#], Elena Nieto Hernández [1,2], Carlo Ceresa[1], Georgios Provatas[3], Karla Ivanković Nizić[3], Milko Jaksic[3], Ettore Vittone[1,2,*], Sviatoslav Ditalia Tchernij[1,2]

[1] Physics Department, University of Torino, via P. Giuria 1, 10125 Torino, Italy

[2] Istituto Nazionale di Fisica Nucleare Sez. Torino, Italy

[3] Division of Experimental Physics, Ruđer Bošković Institute, Bijenička cesta 54, 10000 Zagreb, Croatia

* Corresponding author: ettore.vittone@unito.it

# These authors contributed equally to this work

## Abstract

The Ion Beam Induced Charge (IBIC) technique is widely used to characterize the electronic properties of semiconductor materials and devices. Its main advantage over other charge collection microscopies stems in the use of MeV ion probes, which provide both measurable induced charge signals from single ions, and high spatial resolution, which is maintained along the ion range. It is a fact, however, that the use of low-energy ions in the keV range can provide the IBIC technique with complementary analytical capabilities, that are not available with MeV ions, for example the higher sensitivity to the status, contamination and morphology of the surface and the fact that the induced signal depends on the transport of only one type of charge carrier.

This paper outlines the upgrade that was made at the 100 kV ion implanter of the University of Torino, originally installed for material and surface modification, to explore the rather unexplored keV-IBIC field and to assess its potential to characterize semiconductor devices.

Finally, we report the first IBIC application of our apparatus, which regards the assessment of the radiation damage of a commercially available silicon photodiode, adopting the IAEA experimental protocol and the relevant interpretative model.

## 1. Introduction

Techniques based on the use of accelerators to implant ions with energies in the keV range are widely used for material and surface modification, are the dominant method of doping semiconductors for integrated circuit processing [1,2] and have the potential to significantly impact on the development of physical systems proposed for quantum technologies [3].

More restricted is the use of traditional keV accelerators to analyse materials and devices for their functional, and in particular electronic, characterization, as in the case of the Ion Beam Induced Charge (IBIC) technique, which has been extensively used since early 1990's for the characterization of semiconductor devices, as radiation detectors, high power transistors, solar cells, in conjunction with the study of their radiation hardness and Single Event Upset imaging [4]. Actually, the great majority of IBIC studies are carried out using light (typically H and He) focused MeV ions, since the beam tends to stay "focused" through many micrometers of materials allowing high spatial resolution analysis in thick layers and in buried structures. There are very few applications of IBIC with sub-MeV ion beams [5], which are mainly addressed to the detection of single dopants in silicon [3,6] or diamond [7] using focused sub 30-keV ion beams (FIBs). However, there are at least two reasons, which make "conventional" ion implanters attractive for functional characterization of semiconductors: for example, the high sensitivity of low penetrating ion probes to the status and contamination of the semiconductor surface or the possibility to select only one charge carrier as the responsible of the formation of the induced signal.

In order to explore the analytical potential of the keV-IBIC technique, the Laboratory for Ion implantation at the University of Torino ("LIUTo"), mainly aimed to implant negative single charge ions in materials of interest for quantum technologies [8–10], has been recently upgraded to perform keV-IBIC experiments.

In this paper, we describe the novel IBIC set up at "LIUTo", which uses (max) 100 keV proton collimated beams; IBIC maps were obtained by the synchronous measurement of the photodiode position, which raster scans with respect to the fixed collimator, and of the induced pulse signals, which are processed by a low-noise charge sensitive amplification/acquisition system.

To assess the main spectroscopic and microscopic features of the new setup, a commercially available photodiode was analysed and the spectral and spatial resolutions were extracted from the resulting IBIC spectra and maps, respectively.

Finally, we report on the first IBIC application of our apparatus, which regards the assessment of the radiation damage of a photodiode, previously irradiated with 4 MeV $C^{3+}$ at the Laboratory for Ion Beam Interaction of the Ruder Boskovic Institute of Zagreb, adopting the IAEA experimental protocol and the relevant interpretative model [11].

## 2. Experimental

### 2.1 The ion implanter at LIUTo

The ion implanter at LIUTo is shown schematically in Fig. 1.

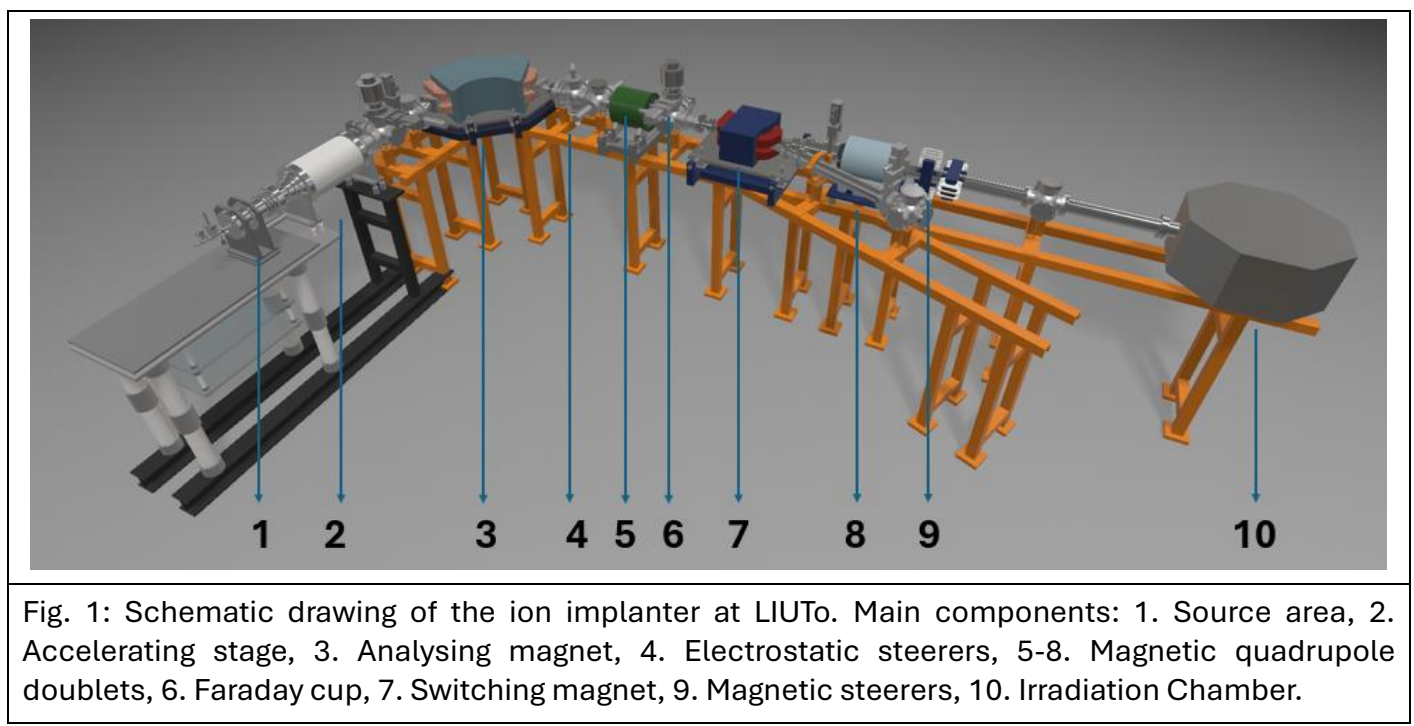

Fig. 1: Schematic drawing of the ion implanter at LIUTo. Main components: 1. Source area, 2. Accelerating stage, 3. Analysing magnet, 4. Electrostatic steerers, 5-8. Magnetic quadrupole doublets, 6. Faraday cup, 7. Switching magnet, 9. Magnetic steerers, 10. Irradiation Chamber.

The Source of Negative Ions by Cesium Sputtering (model SNICS II from National Electrostatics Corp. [12]) was chosen due to its versatility to produce many different negative ion species [13], which is a basic requirement for color center production in semiconductors and insulators [8,9].

Negative charged sputtered atomic or molecular ions are produced by the sputtering process induced by Cs+ ions accelerated to few keV of energy towards the cathode target, and by the subsequent capture of electrons by Cs vapor or by the Cs layer deposited on the cathode surface. Negative ions passing through the extracting electrodes are then injected into the accelerator stage (NEC model 2FA011060, powered by a high-voltage negative stabilized power supply Heinzinger PNC 100000-3), reaching a maximum energy of about 120 keV (resulting from the sum of the cathode voltage, extractor voltage and high voltage).

For electrical and radiation safety [14], the source and the high voltage area are located in a Faraday cage with interlocked access door. Since the highest amount of ionizing radiation (electron and x-rays) is generated in proximity of the source and the beam extraction region, the Faraday cage walls are covered with 2 mm thick lead panels. The lead thickness was chosen in order to keep workers' exposure below the effective dose values to classify as category B the exposed workers (EU Directive 59/2013 [15]).

The other main components (numbers in the round brackets refer to the legend of Fig. 1) of the ion implanter are the following:

a. The 90° magnetic mass analyzer (3) is a water-cooled magnetic dipole with bending radius of 500 mm and maximum field intensity of 0.83 T (from Bruker).
b. Two (5,8) magnetic quadruple doublets (Varian mod. 1050) for beam focusing.
c. Custom made electrostatic (4) and magnetic (9) steerers, used for beam alignment in the transverse (X–Y) plane. The second beam line, at 21° against the beam axis, accessible through the switching magnet (7) (from Sigma Phi), is currently not operative.
d. One cooled fixed aperture (diameter = 20 mm) and a motorized set of slits are located upstream and downstream the 90° analyzing magnet, respectively.
e. A custom-made spring-loaded Faraday cup (aperture= 2 cm) (6) with electron suppression assembly to monitor the ion beam current downstream the magnet.
f. Four gate valves (VAT CF100) fraction the beam line in three parts (source, analyzer and switching magnets); each of them is served by an oil-free pumping system (Turbovac 350i backed by a Scrollvac 7 plus from Leybold), which maintain the residual gas pressure down to less than $10^{-4}$ Pa. The fourth gate valve isolates the beam line from the target chamber, equipped with a similar independent pumping system.
g. The implanter at LIUTo uses USB NI6001 cards from National Instruments controlled by the Labview programming environment to interface the beam-line main components (i.e. vacuum handling, slit motion).

Finally, the custom designed target chamber (10) is hosted in a class 10000 clean room. The observation of the beam size and of its position is made by a CsI(Tl) scintillator located at the target position, whose iono-luminescence is imaged by a CMOS camera (CS165 Zelux from Thorlab). A Keithley 6485 pico-amperometer measures the ion beam current collected by a custom-made Faraday cup (aperture of 1.5 mm diameter) with electron suppressor. The target chamber is equipped with five micro positioners. The first two consist of closed-loop piezo-walk axes (Phisik Instrumente model N-565K015) and are used to translate the collimating aperture, which faces the target, with nanometric resolution and repeatability along the x-y plane (perpendicular to the beam). The target is mounted on a three-axis micro-positioning system (Phisik Instrumente mod. 6230V6400 and 6230V1000 for the xy plane and model M-112.1VG1 for the z direction).

For this IBIC experiment, the experimental set-up is schematically shown in Fig. 2. The collimator consists of a ≈ 4 µm diameter aperture micromachined by laser milling in a 50 µm thick steel plate and mounted on the first two micro-positioners. The device under test (DUT) is mounted on the second set of positioners, at a distance of ≈ 0.3 mm from the collimator.

The proton beam was formed using the parameters listed in Table 1.

| Cathode | Titanium hydride |
|---|---:|
| Cathode potential (0:-15 keV) | -5 kV |
| Extractor potential (0:15 kV) | 6 kV |
| Accelerator potential (0-100 kV) | 89 kV |
| Table 1: Source parameters used for the IBIC experiment described in this work | |

To locate the beam spot, the ion beam was focused down to about 1 mm spot size at the target position. The beam was then defocused to achieve the lowest measurable current by the Faraday cup (fractions of pA) in the irradiation chamber and both the collimating aperture and the DUT were positioned at the beam focus. Operating conditions are met when the DUT's pulsed signal frequency is in the 1000 counts per seconds (cps) range. In such a configuration, IBIC maps are acquired by scanning the photodiode with respect to the ion beam, which is collimated by the fixed mask.

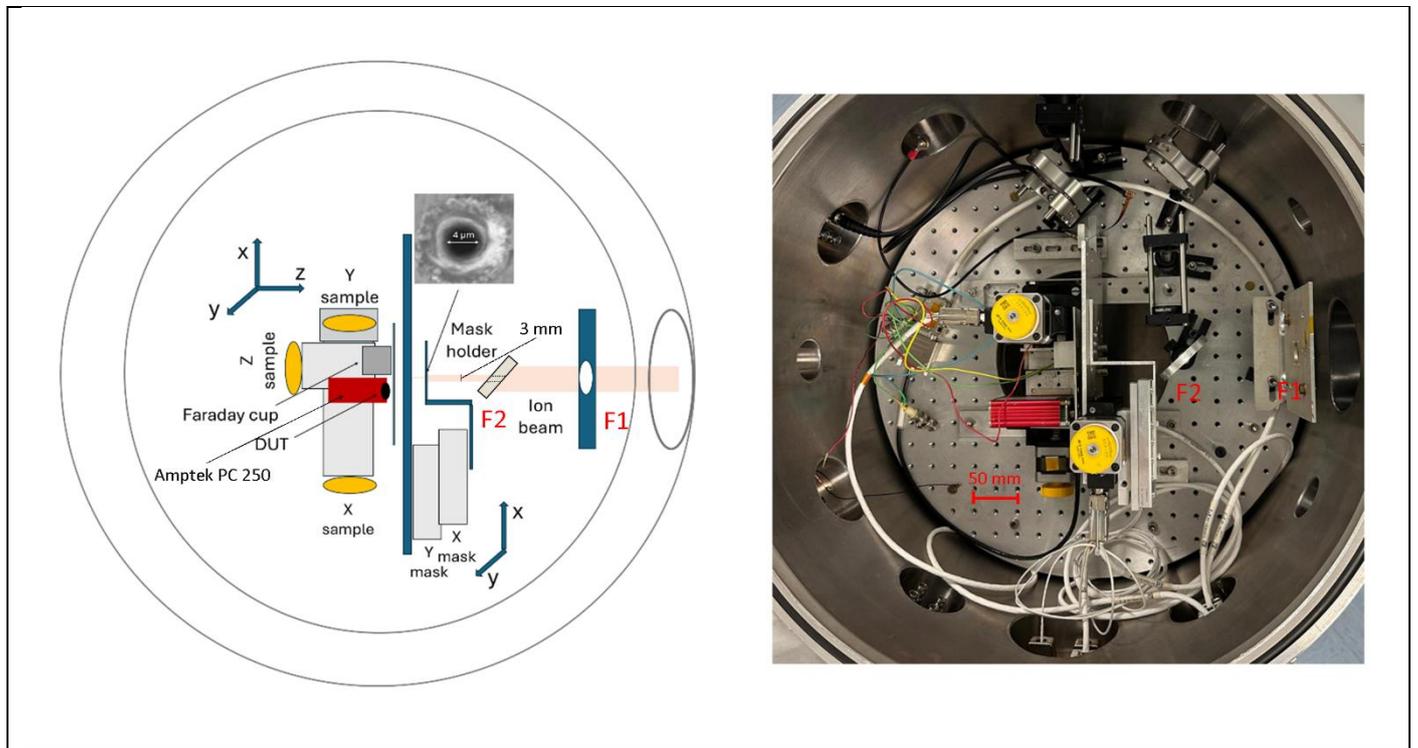

Fig. 2: Schematic representation of the experimental set-up inside the irradiation chamber (left) and photo of the setup (right). The beam enters the chamber through a first collimating aperture (F1) necessary to shield the instrumentation from stray ions, a through-hole mirror (F2 - 3 mm diameter aperture) is present for the optical imaging of the samples and collimation masks. The collimation masks and the sample holder are mounted together on the same mechanical support in order to keep

the reference fixed for both the coordinate systems. A Faraday cup and the DUT on a PC 250 test board by Amptek (enclosed in the red aluminum box in figure) are both in-built on the sample holder.

## 2.2 Scanning and data acquisition system

The schematic of the scanning and of the data acquisition system is shown in Fig. 3.

The photodiode was connected in close proximity to the Amptek A250 pre-amplification circuit [16]. The resulting signal was then amplified and filtered by Silena model 7611 shaping (shaping time 1 µs) amplifier.

The unipolar output signal was sent to the multichannel analyser (Ortec Easy MCA) and to the oscilloscope (Teledyne LeCroy Wave Surfer). The former was used to acquire the energy spectra, while the latter was employed for IBIC mapping. One channel of the oscilloscope was dedicated to sample the IBIC signal, providing 8-bit digitization within the range of 50 mV to 450 mV. The other three channels were used to acquire a veto signal and to identify the x-y coordinate of the scan on the basis of two analog values in the range 0V and 10V.

The scheme for IBIC map acquisitions relies on two different computers, (labelled as PC 1 and PC 2). PC 2 was used to control the scanning software and to send to the oscilloscope three different signals via a National Instruments card (Model 6343): i) logic signal (labelled as veto in Fig. 3), ii-iii) 0 to 10 V signal. The NI card allowed to measure the rate of ions impinging on the photodiode over time. PC1 was used to interface to the oscilloscope and to convolute the coordinates of the positioners with respect to the acquired IBIC signal and hence to reconstruct the IBIC map of the irradiated areas.

The veto signal had the purpose of selecting the measurement window and avoiding the acquisition of noisy data, which was originated by the DUT movement. The x(/y) reference step consisted of an analog voltage signal ranging from 0 to 10 V depending on the adopted resolution (minimum step) and ranges for the scan.

The acquisition was triggered by an IBIC signal crossing a defined threshold at the oscilloscope, the software retrieved from the oscilloscope the average value acquired on the veto and position channels, along with the peak value from the IBIC signal channel. If the veto signal was low, the data was immediately discarded. If high, the position (hence voltage reference from the NI card) and signal values from the photodiode were stored. The veto signal was set low when the system is in motion and raised during the acquisition window.

The IBIC maps were acquired with a minimum step below the size of the aperture, i.e. < 4 µm.

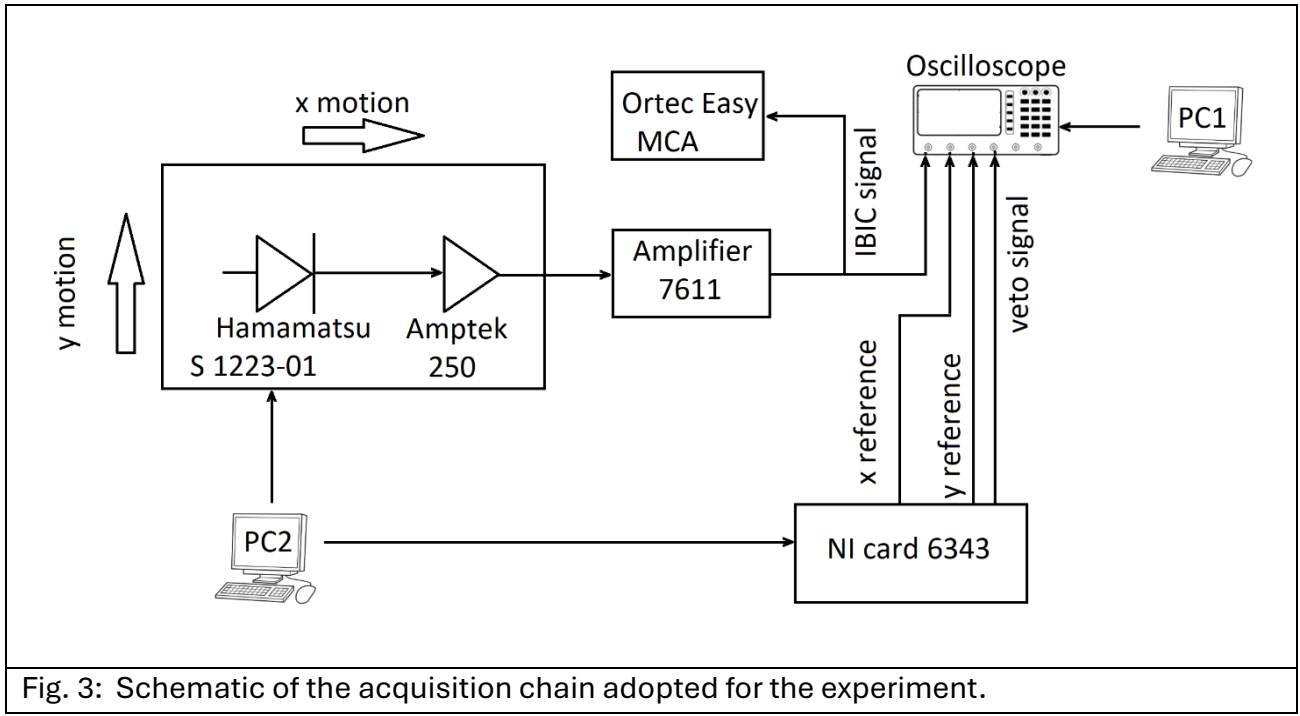

Fig. 3: Schematic of the acquisition chain adopted for the experiment.

## 2. Results

To assess the spectroscopic and microscopic features of the experimental set-up described in the previous section, we used the IBIC technique to characterize a commercial silicon p-i-n photo-diode, Hamamatsu S1223-01 (DUT), whose metal can package was opened to remove the borosilicate glass window, to expose the frontal p-type surface to the ion beams [11].

This type of photodiode was extensively characterized in a previous work [17] and its main features are here summarised: donor concentration of the n-type layer is about $7 \cdot 10^{12}$ cm$^{-3}$, the p-type top layer extends down to 0.6 µm from the surface with a maximum acceptor concentration of $2 \cdot 10^{19}$ cm$^{-3}$. The surface is covered with 110 nm of silicon oxide as measured by RBS.

The IBIC measurements were carried out using collimated proton beams of energy 40-100 keV, in dark conditions and at room temperature.

The charge pulse rate was of the order of 1000 cps and the bias voltage was set at 25 V, which we found to be the optimal choice to guarantee the full collection of the induced charge (the peak position remained constant from 25 to 85 V) and to minimize the noise (the spectral FWHM linear increases in the bias voltage range 25-85 V).

### 3.1 Spectroscopic characterization

To calibrate the electronic chain, IBIC spectra were acquired using protons of energy ranging from 40 to 100 keV (Fig. 4a). By plotting the proton energy as function of the signal amplitude (expressed in channel number) peak spectrum (Fig. 4b), the sensitivity of the electronic chain can be extracted from the slope of the linear fit (657±9) eV/channel, corresponding to about 183 electrons/channel, assuming an average energy of 3.6 eV to create an electron–hole pair in silicon [18].

The mean FWHM calculated from the gaussian fits of the IBIC spectra (Fig. 4a) was (9.4±0.3) channel, which correspond to (6.2±0.2) keV (1720 electrons). This value is the effective resolution of the IBIC set-up, being the convolution of the contribution of the electronic chain and of the broadening of the energy spectrum of ions incident on the DUT, due to the scattering of ions with the collimator edges.

The intercept of the linear fit shown in Fig. 4b is (20.4±0.7) keV. Assuming an average (with respect to all the energies) ionization stopping power of protons in Si of (121±4) eV/nm [19] (the uncertainty is given by the dispersion of the stopping power data for the proton energy range 40-100 keV), the effective dead layer thickness is (169±8) nm, in good agreement with a previous estimate (180 nm) based on Angle Resolved IBIC measurements carried out with 2.38 MeV Li ions [17]. Considering the top oxide layer (110 nm thick), the thickness of silicon beneath the electrode ($p^+$ layer), in which charge collection is inefficient is 60-70 nm.

The noise threshold was fixed to channel 14, which, according with the previous calibration, corresponds to about 30 keV.

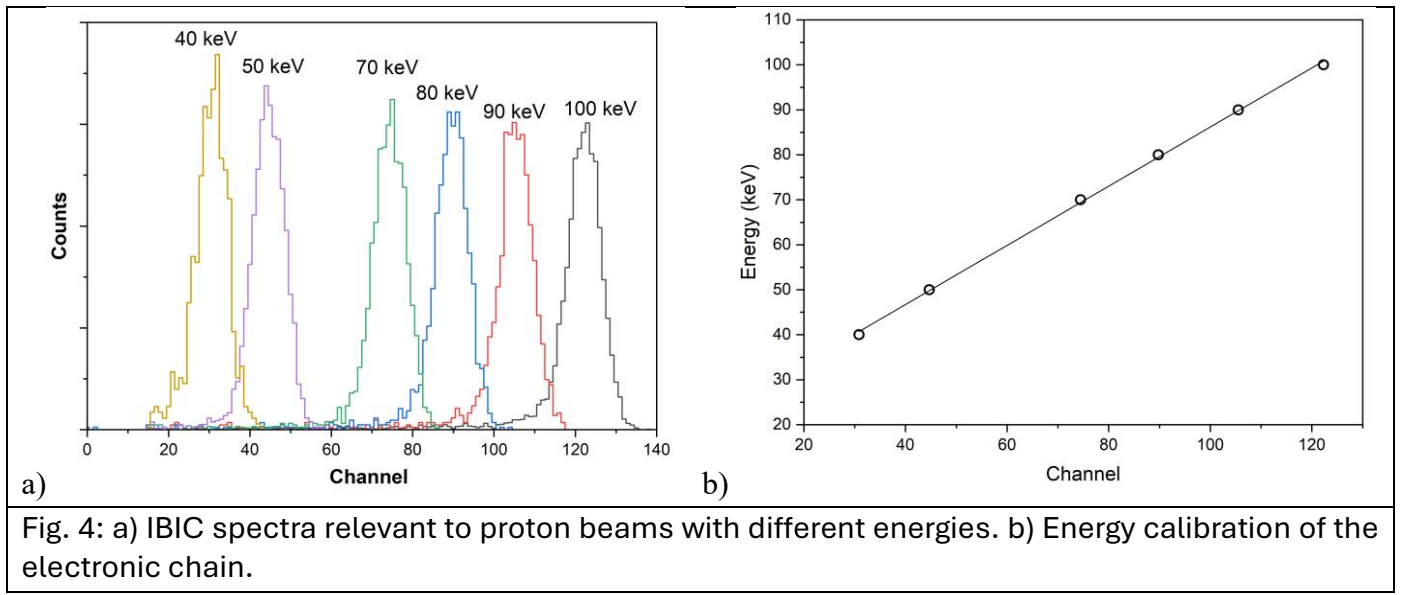

Fig. 4: a) IBIC spectra relevant to proton beams with different energies. b) Energy calibration of the electronic chain.

### 3.2 Microscopic characterization

To assess the spatial resolution of the apparatus, we exploited the contrast of the IBIC map relevant to a region around the DUT strip electrode shown in Fig. 5a.

The DUT was scanned (step size = 3 µm) with respect to the fixed collimator to cover a 100x100 µm$^2$ region centred around the strip electrode. The pulses generated by the collimated 100 keV proton beam were acquired and stored by the data acquisition system in the list mode ("event by event" [11]), along with the coordinates of the DUT.

Fig. 5b shows the map of the median pulse height stored at each step. It is apparent that in proximity of the electrode, the collection efficiency decreased, as clearly shown in Fig. 6, which represents the median amplitude (the error bar is the standard deviation) of pixels of the IBIC map (Fig. 5b) equidistant from the reference line shown in the lower left corner and parallel to the electrode main axis.

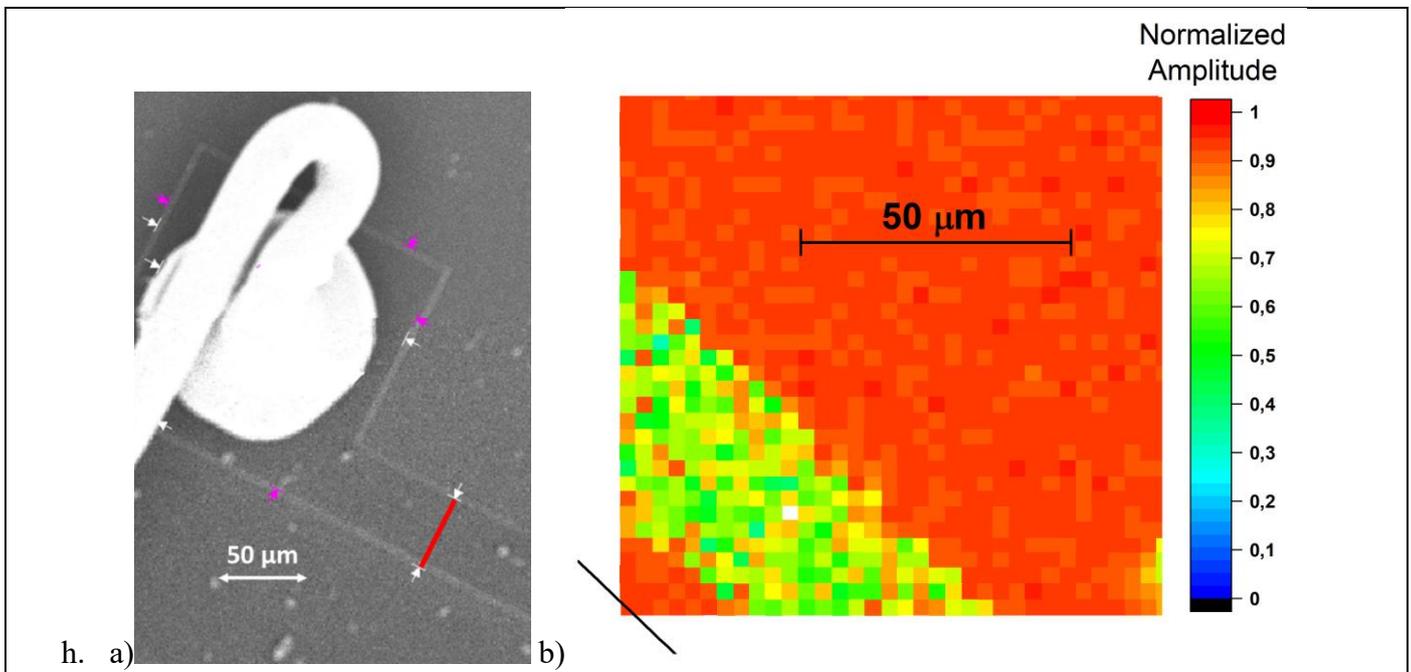

Fig. 5: a) Scanning electron image of the DUT surface, with the wire bonding and the electrode. The red segment is 47 μm wide. b) IBIC map of a 100x100 μm$^2$ region around the strip electrode. The colors represent the median of the pulse amplitude acquired for each DUT position. The color scale on the right encodes the median amplitude which is normalized to the signal emerging from the central uniform region. The segment at the lower left corner is the reference line used to calculate the amplitude profile shown in Fig. 6.

The regression of the data against error/complementary error functions was performed to quantify the spatial resolution, which was (7.6 ± 0.7) μm, determined as the FWHM of the first derivative of the fitting

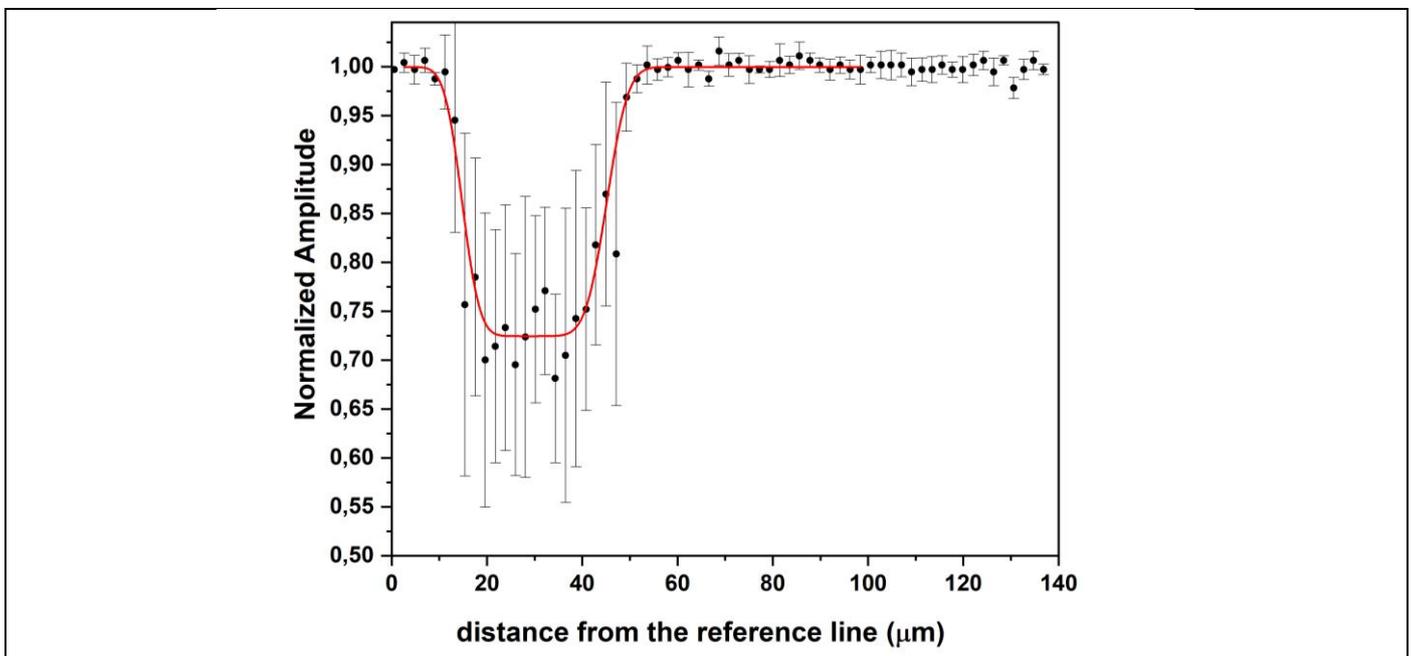

Fig. 6: amplitude profile along the direction orthogonal to the reference line shown in Fig. 5b. The markers and the error bars represent the median and the standard deviation of the signal amplitudes, respectively. The red curve is the representation of the combination of the error and complementary error functions fitting the experimental data.

curve.

## 3.3 Radiation hardness assessment

The first IBIC experiment conducted at LIUTo was aimed to assess the radiation hardness of the DUT, adopting the protocol developed within an IAEA Coordinated Research Project (CRP, reference F11016) "Utilization of Ion Accelerators for Studying and Modelling Ion Induced Radiation Defects in Semiconductors and Insulators" [11,20].

As a first step, nine non overlapping regions (about $10^4$ µm$^2$) of the DUT were irradiated at the microprobe facility of the Ruder Boskovic Institute [21] by raster scanning a 4 MeV $C^{3+}$ ion beam focused down to 5 µm spot size with the fluences listed in Table 2.

| Region | 0 | 1 | 2 | 3 | 4 | 5 | 6 | 7 | 8 | 9 |
|---|---|---|---|---|---|---|---|---|---|---|
| Fluence ($10^{10}$ ions/cm$^2$) | 0 | 1.15 | 3.23 | 5.38 | 7.53 | 10.8 | 32.3 | 54.6 | 76.1 | 96.1 |

Table 2: 4 MeV $C^{3+}$ ion fluences of the 9 irradiated regions.

After the irradiation, the reverse current increased from few nA (in the pristine case) to 105 nA for a bias potential of 23 V.

After a few days of irradiation, during which the samples were kept at room temperature, the irradiated regions were probed using a 100 keV proton collimated beam at LIUTo using the same experimental procedure described in section 2,2. The IBIC measurements were carried out by applying a bias voltage of 25 V. However, due to the potential drop across the load resistance (213 MΩ), the effective polarization of the DUT was about 2 V.

The resulting IBIC map (Fig. 7a) clearly shows the degradation of the charge induced signal as the 4 MeV $C^{3+}$ ion fluence increases. This is furtherly evidenced in Fig. 7b, which shows the pulse spectra extracted from the central region (50x50 µm$^2$) of the irradiated areas (0-7). The spectrum relevant to the pristine region "0" has a FWHM corresponding to 12 keV, which is approximately twice the FWHM measured in the pristine diode (see section 2.1) due to the remarkable increase of the noise induced by the high leakage current. Pulse spectra relevant to regions 8 and 9 are not reported since they were remarkably affected by noise.

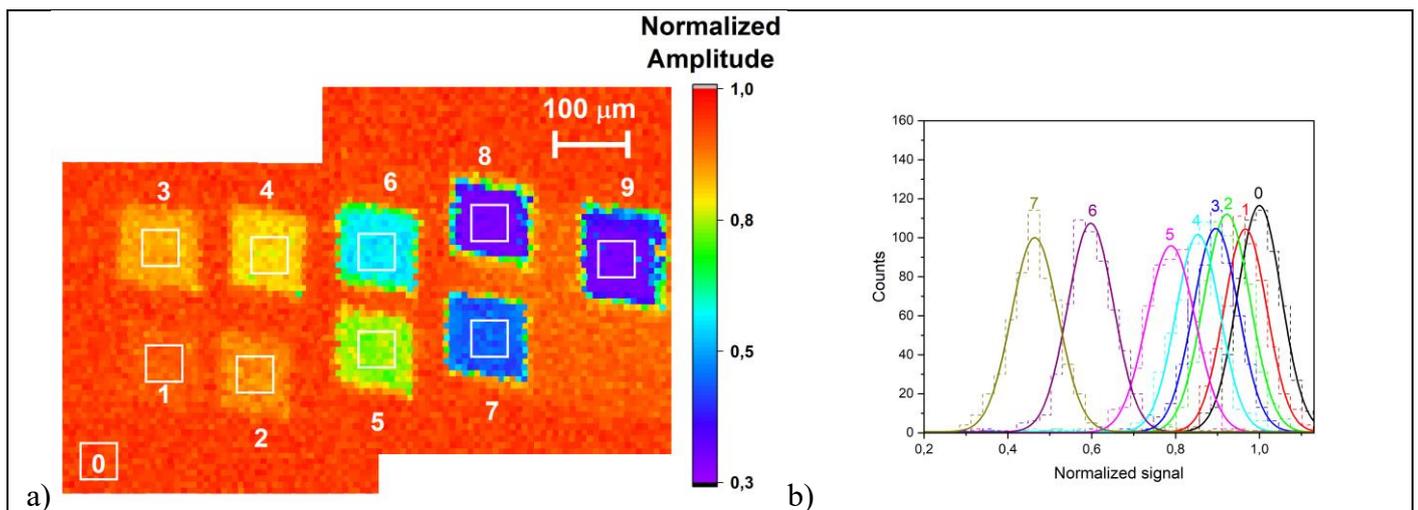

Fig. 7: a) IBIC map of the nine regions irradiated with 4 MeV $C^{3+}$ ions with fluences listed in Table 2. The white squares are the extension (50x50 µm$^2$) of the regions from which the damaged and undamaged CCE spectra are extracted. The spectrum of the pristine region was extracted from area labelled "0". The colour scale on the right side encodes the median pulse for each DUT position normalized to the

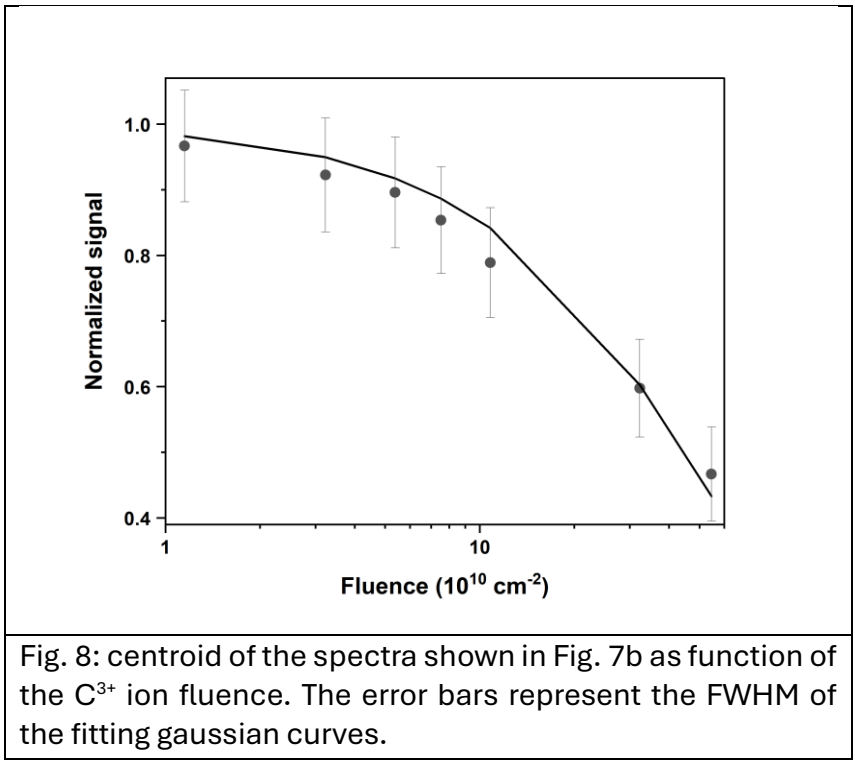

Fig. 8: centroid of the spectra shown in Fig. 7b as function of the $C^{3+}$ ion fluence. The error bars represent the FWHM of the fitting gaussian curves.

Fig. 8 shows the centroids of the fitted spectra as a function of the damaging (4 MeV $C^{3+}$) ion fluences. The analysis of these results was carried out through the model described in the above mentioned IAEA protocol [11,20].

The fitting procedure, detailed in the supplementary material, provides results represented by the continuous line in Fig. 8; the resulting capture coefficient for electrons is

$\alpha_n$=(3300±500)$\mu m^3$/s,

This value is significantly higher than the value obtained using MeV He ions both to damage and probe the charge collection induced degradation of a similar diode [17]. To explain this discrepancy, it is worth noticing that the present result refers only to electrons, whereas in [17], the charge collection efficiency (CCE) degradation is due to the combination effects of both the carriers (electrons and holes), and the relevant capture coefficients were found to be strongly correlated. Moreover, the model adopted to fit the experimental data assumes that the recombination centres produced by ion induced damage are solely due to vacancy production, which is a reasonably assumption when using He ions. Using C ions, our result suggest a higher probability of divacancy production evaluated by SRIM code or an increasing importance of Si-C complexes.

## 4. Conclusions

In this paper we presented the application of the IBIC technique at the "LIUTo" ion implanter, which was originally designed and commissioned to implant negative ions in materials of interest for quantum technologies.

The IBIC experiment was carried out on a commercial silicon photodiode, which was mounted on a three-axis motorized stage for precise alignment with the proton beam emerging from a collimator with a central aperture of 4 µm, placed at a distance of about 0.3 mm from the sample.

The low-noise acquisition system was calibrated using proton beams in the energy range of 40-100 keV, whose pulse spectra show a spectral resolution of about 6 keV. The intercept of the calibration curve provides an estimate of the dead layer thickness, which is in complete agreement with previous Angle Resolved IBIC and RBS measurements [17].

The spatial resolution, about 8 µm, was measured by scanning a region centered around a strip electrode of the pristine photodiode with respect to the fixed collimated proton beam and synchronously acquiring the pulse height and the sample position, mainly resulting from the scattering of protons from the collimator aperture edges.

The first application of the IBIC set-up was devoted to assess the radiation hardness of the photodiode, following the IAEA guidelines [11]. The resulting capture coefficient regards only electrons and is not affected by the correlation with the analogous coefficients relevant to electrons and holes, resulting from previous measurements carried out using more penetrating MeV He ions.

Although our system is less efficient than traditional systems using focused MeV or keV ions, there are rooms of improvements in the spatial resolution (e.g. using micro [22] or nano [23]-capillary) and in the acquisition system. However, at the state of the art, LIUTo is proven to be suitable to perform IBIC measurements and pave the way to the use of similar set-ups for applications requiring low penetrating ions, i.e. in which only one carrier induce the charge at the sensing electrode and for precise analysis of surface effects.

## Acknowledgements

This work was financially supported in the context of the following projects: PNRR MUR project PE0000023-NQSTI; experiments ROUGE and Quantep funded by the 5th National Commission of the Italian National Institute for Nuclear Physics (INFN); the 'Training on LASer fabrication and ION implantation of DEFects as quantum emitters' (LasIonDef) project, funded by the European Research Council under the 'Marie Skłodowska-Curie Innovative Training Networks' program, Departments of Excellence" (L. 232/2016), funded by the Italian Ministry of Education, University and Research (MIUR). The authors acknowledge prof. J. Meijer and his group of the University of Leipzig (D) for the design and the commissioning of the ion implanter. E.V., S.D.T, A.B. and E.N.H would like to thank Dr. Natko Skukan of the Nuclear Science and Instrumentation Lab. of the International Atomic Energy Agency (IAEA) for his valuable help.

## Supplementary material

The fitting procedure adopted to analyse the Charge Collection Efficiency degradation shown in Fig. 8 is based on the model described in detail in [11] and applied in [17,20]. It is based on the Shockley-Ramo-Gunn theory to model the induced charge formation mechanism and on the Shockley-Read-Hall model for the free carrier recombination in the presence of ion-induced deep traps.

The experiment described in Section 3.3 fulfil the assumptions listed in Appendix I of [11,20]. In particular, the depletion region extends well beyond the range of the damaging 4 MeV $C^{3+}$ and probing 100 keV $H^-$ ions, which are 4.5 µm and 1.0 µm, respectively, as calculated by SRIM computer code (Fig.S 1a).

For this specific experiment, the general expression [17,20] can be simplified, considering only the contribution of majority carriers (electrons), since the generation volume extends to less than 1 µm from the anode.

Therefore, the normalized charge collection efficiency (CCE) degradation reduces to:

$$\text{CCE} = \int_0^d dx \left\{ \gamma(x) \cdot \int_x^d dy \left\{ \mathcal{E}^+(y) \cdot \exp\left[ -\alpha_e \cdot \Phi \cdot \int_x^y dz \frac{\eta(z)}{v_e(z)} \right] \right\} \right\}$$

Where

$\gamma(x) = \frac{1}{E}\frac{dE}{dx}$ is the SRIM simulated normalized ionization energy loss profile relevant to 100 keV protons; E is the ion energy (Fig.S 1a)

$\eta(x)$ is the vacancy profile (the number of vacancies generated by 4 MeV $C^{3+}$ per ion per unit length) (Fig.S 1a) as simulated by SRIM assuming a displacement energy of 21 eV[1]

$v_e$ is the electron drift velocity profile relevant to an applied bias voltage of 2 V, calculated using the SILVACO device simulator[2] (Fig.S 1b)

$\mathcal{E}^+(z) = \frac{\partial \mathcal{E}(V_{bias},z)}{\partial V_{bias}} = \frac{\mathcal{E}(V_{bias}+\Delta V,z) - \mathcal{E}(V_{bias}-\Delta V,z)}{2\Delta V}$ is the Gunn's weighting field profile (Fig.S 1b); $\mathcal{E}(V,z)$ is the electric field calculated using the SILVACO device simulator[2] relevant to an applied bias voltage $V_{bias}$=2V. for numerical differentiation, the derivative was evaluated using a finite difference approximation, with an increment $\Delta V = 0.01\ V$.

---

[1] J.W. Corbett, G.D. Watkins, Production of divacancies and vacancies by electron irradiation of silicon, Phys. Rev. 138 (1965). https://doi.org/10.1103/PhysRev.138.A555.

[2] https://silvaco.com/tcad/

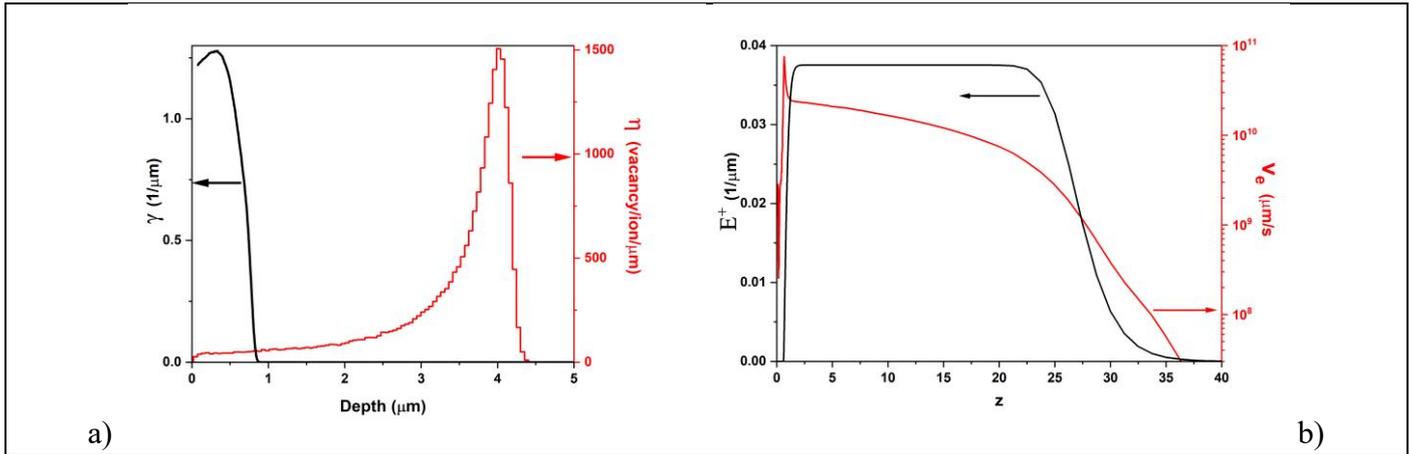

Fig.S 1: : a) Normalized ionization energy loss ($\gamma$) of 100 keV protons in Si (left scale, black curve) and Vacancy profile ($\eta$) for 4 MeV C ions in Si (right scale, red curve) from SRIM simulation; b) electron drift velocity $v_e$ (right scale, red curve) and Gunn's weighting field $\mathcal{E}^+$ (left scale, black curve)